# Dielectric Fano Nanoantennas for Enabling Sub-Nanosecond Lifetimes in NV-based Single Photon Emitters

Shu An[1], Dmitry Kalashnikov[1], Wenqiao Shi[2], Zackaria Mahfoud[1], Ah Bian Chew[1], Yan Liu[1], Jing Wu[1,3], Di Zhu[1,4], Weibo Gao[5], Cheng-Wei Qiu[2,*], Victor Leong[1,*], and Zhaogang Dong[1,4,*]

[1]Institute of Materials Research and Engineering, A*STAR (Agency for Science, Technology and Research), 2 Fusionopolis Way, #08-03 Innovis, 138634 Singapore

[2]Department of Electrical and Computer Engineering, National University of Singapore, 4 Engineering Drive 3, Singapore 117583, Singapore

[3]School of Electronic Science & Engineering, Southeast University, Nanjing, 211189, China

[4]Department of Materials Science and Engineering, National University of Singapore, 9 Engineering Drive 1, 117575, Singapore

[5]Division of Physics and Applied Physics, School of Physical and Mathematical Sciences, Nanyang Technological University, Singapore 637371, Singapore

*Correspondence and requests for materials should be addressed to Z.D. (email: dongz@imre.a-star.edu.sg), V.L. (email: Victor_Leong@imre.a-star.edu.sg), and C.-W.Q. (email: chengwei.qiu@nus.edu.sg).

**ABSTRACT**

Solid-state quantum emitters are essential sources of single photons, and enhancing their emission rates is of paramount importance for applications in quantum communications, computing and metrology. One approach is to couple quantum emitters with resonant photonic nanostructures, where the emission rate is enhanced due to the Purcell effect. Dielectric nanoantennas are promising as they provide strong emission enhancement compared to plasmonic ones, which suffer from high Ohmic loss. Here, we designed and fabricated a dielectric Fano resonator based on a pair of silicon (Si) ellipses and a disk, which supports the mode hybridization between quasi-bound-states-in-the-continuum (quasi-BIC) and Mie resonance. We demonstrated the performance of the developed resonant system by interfacing it with single photon emitters (SPEs) based on nitrogen vacancy ($NV^-$) centers in nanodiamonds (NDs). We observed that the interfaced emitters have a Purcell enhancement factor of ~10, with sub-ns emission lifetime and a polarization contrast of 9. Our results indicate a promising method for developing efficient and compact single-photon sources for integrated quantum photonics applications.

**Introduction**

Bright single photon emitters (SPEs) are essential elements for quantum communications, quantum-enhanced metrology and quantum computing[1, 2]. While many efforts are dedicated to the development of the new bright SPEs, another approach relies upon improving emission properties of the existing ones by interfacing them with resonant photonic structures and benefiting in increased brightness of emitters due to the Purcell effect[3]. Moreover, this approach provides additional control over spatial and polarization degrees of freedom of the emission. Many successful attempts have already been demonstrated with plasmonic and hyperbolic metamaterials, and viable dielectric thin film platforms (*i.e.*, $TiO_2$, AlN, $Si_3N_4$)[4-11].

Different platforms have their own pros and cons. For example, plasmonic nanoantennas provide significant field enhancement with demonstrated record level lifetime reduction[12-14]. However, their high optical absorption brings a significant non-radiative contribution to the total decay rate. On the contrary, dielectric nanoantennas are free from such strong non-radiative contributions while maintain the advantage of tunability of resonance wavelength and model structure[6, 15, 16]. Additionally, some of the dielectric materials (Si, $Si_3N_4$) are CMOS-compatible, which is important for the potential large-scale fabrication. In addition, hyperbolic metamaterials benefit in broad range of supported wavelengths, and some remarkable results with photoluminescence enhancement and lifetime reduction have been demonstrated[17, 18]. Nevertheless, their interfacing with SPEs is still very limited even with sophisticated fabrication process[18, 19].

In this work, we designed and fabricated all-dielectric silicon metasurface that has a sharp Fano resonance at the visible range to achieve strong Purcell enhancement factor. We tested the performance of the designed Fano antenna with nitrogen vacancy ($NV^-$) centers in nanodiamonds (NDs), one of the fluorescent atomic defects, which are commonly served as SPEs due to its remarkable optical properties as well as the convenience and scalability of solid-state host system[11, 20, 21]. They are stable under room temperature operation, facilitating the rapid characterization and therefore development of practical quantum communication systems[22]. However, they suffer from poor polarization selection, low brightness and long fluorescence lifetime[23, 24], which is typically around 10 to 30 ns[25, 26]. Thus, its limited emission rate poses a challenge for their integration into high-speed quantum devices. Correspondingly, significant improvement of their emission properties upon interfacing with our Fano metasurface would mark successful realization of our system as promising candidate for quantum photonics.

Our dielectric Fano resonance were realized by mode hybridization between a pair of Si ellipses supporting quasi-bound-states-in-the-continuum (quasi-BIC) and a Si nanodisk, which supports Mie resonance[27-29]. Numerical simulations were performed to optimize the geometric parameters of the Fano nanoantenna to ensure a good spectrum overlap between its resonance and the emission spectrum of NV$^-$ centers. Afterwards, the NV$^-$ centers were integrated on the metasurface and tested by measuring second-order autocorrelation function $g^{(2)}(0)$, to be quantum emitters (*i.e.*, $g^{(2)}(0) < 1$) and/or SPEs (*i.e.*, $g^{(2)}(0) < 0.5$). Their measured emission lifetimes were reduced significantly down to 0.93 ns, which is one order of magnitude shorter as compared to the ones on flat substrate region. For the best of our knowledge, the achieved Purcell factor is the one of the highest values so far for quantum emitters interfaced with all-dielectric metasurfaces [30, 31]. Additionally, it was proved that our nanoantenna arrays enable the SPE with polarization control. We believe that our findings provide a significant step towards the practical implementation of bright single photon sources for quantum technologies.

**Results and Discussion**

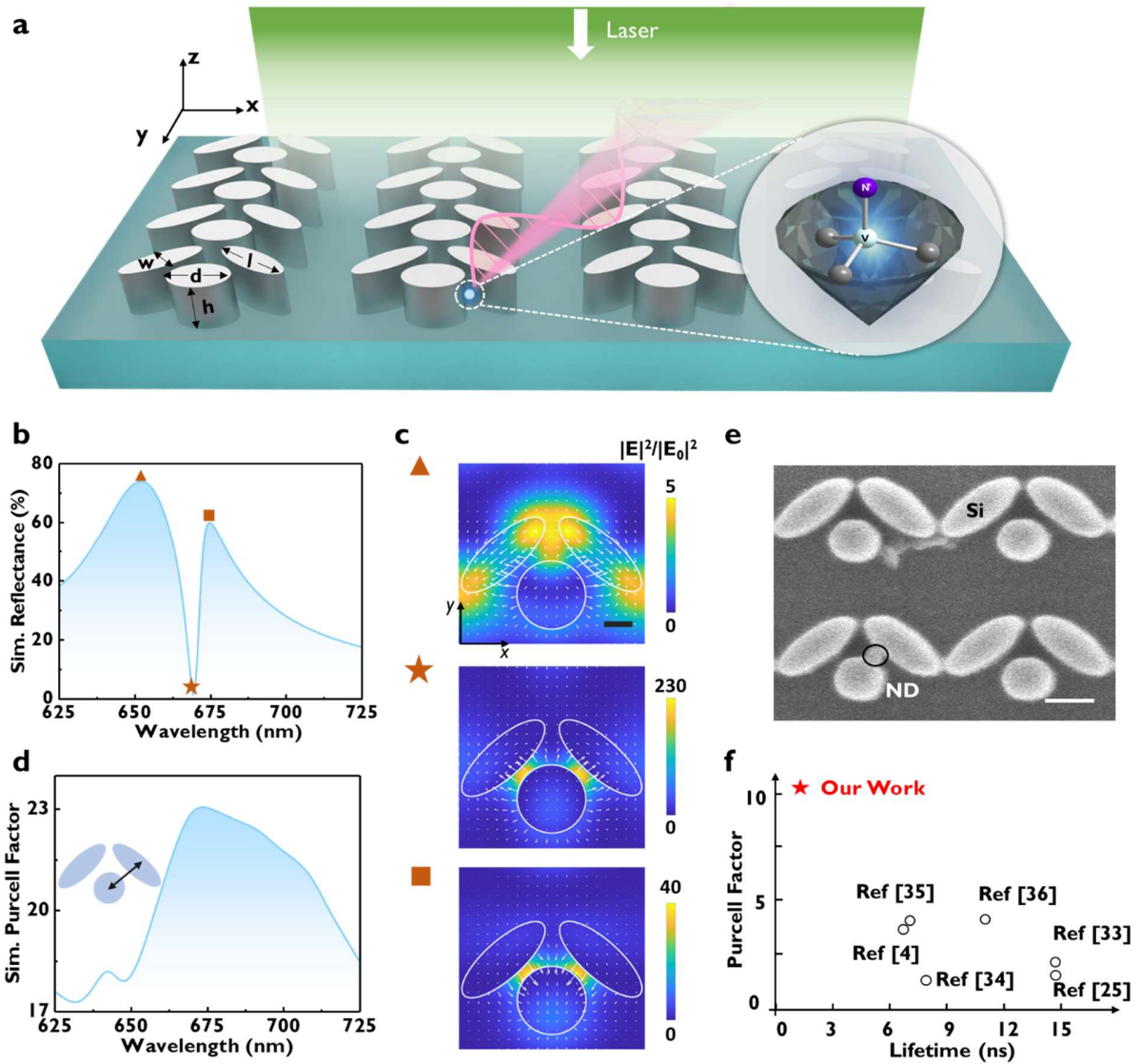

**Fig. 1. Single photon emitter integrated with Si optical nanoantenna exhibiting Fano resonance. a** Schematic of the nanodiamond NV[-] center SPE integrated with nanoantenna array on a crystalline-Si (240 nm) on sapphire substrate. **b** The illustration of Fano resonance mechanism using the simulated reflectance spectrum. **c** Electric field intensity distributions under *y*-polarized incidence light at the respective wavelengths marked by the symbols in **b**. The scale bar is 50 nm. **d** Simulated Purcell factor of the Fano nanoantenna when a dipole source is placed in the nanogap

between ellipse and disks. **e** Scanning electron microscope (SEM) image showing the diamond NV⁻ centers being deposited onto the metasurface with a scale bar of 100 nm. **e** Benchmarking of the achieved emission lifetimes and Purcell factors for diamond NV⁻ center SPEs as coupled to dielectric nanoantennas.

To achieve the bright and polarized quantum light sources with enhanced emission rates via the Purcell effect, we introduce a Si metasurface with Fano resonance composed of a pair of ellipses and middle disk shown in Figure 1a. The designed Si ellipse pair with a length ($l$) of 176 nm and a width ($w$) of 64 nm is rotated 48° along the *x*-axis, thus introducing asymmetry for achieving the quasi-BIC resonances[32]. Moreover, the Si nanodisk with diameter ($d$) of 120 nm is displaced 75 nm in the positive direction of the *y*-axis direction. The height ($h$) of the Si nanostructure was designed to be fixed at 240 nm, with a pitch size of 327 and 324 nm along *x* and *y* axis, respectively. Here, dip-coating method was carried out to deposit diamond nanoparticles with NV⁻ centers onto the metasurface.

To elaborate the mechanism of the Fano resonance, the simulated reflectance spectrum under *y*-polarized incidence condition and the corresponding electric intensity distributions were shown in Figure 1b and Figure 1c, respectively. The quality factor of resonance at 671 nm was evaluated to be 200, while the measured reflectance spectrum is at supporting information (Figure S1). The resonance can be further explained by the electric field intensity changes in Figure 1c and magnetic field distribution (see SI, Figure S2). At the dip of the reflectance spectra in Figure 1b (indicated by star), strongest electric intensity occurs at the nanogaps between disk and two ellipses, also called hotspot, where the maximum intensity ($|E|^2/|E_0|^2$) enhancement factor is 230. When an in-plane electric dipole, oriented as indicated in the black arrow (inset of Figure 1d), is placed in the

nanogap, the simulated Purcell factor is up to 23 times (Figure 1d). The position of ND with NV⁻ center emitter is confirmed by SEM image shown in Figure 1e, where a single ND with a diameter of 35 nm is clearly visible at the nanogap between the disk and ellipses.

Figure 1f presents the benchmarking summary of NV center SPEs in ND on the dielectric antenna being compared with respect to previous experimental works in terms of lifetime and Purcell factor [4, 25, 33-36]. Based on our knowledge, our work outperforms over the other dielectric nanoantenna enabled NV center-based SPEs in terms of emission lifetime, highlighting the significance of achieving high Purcell factor when using the Fano resonance.

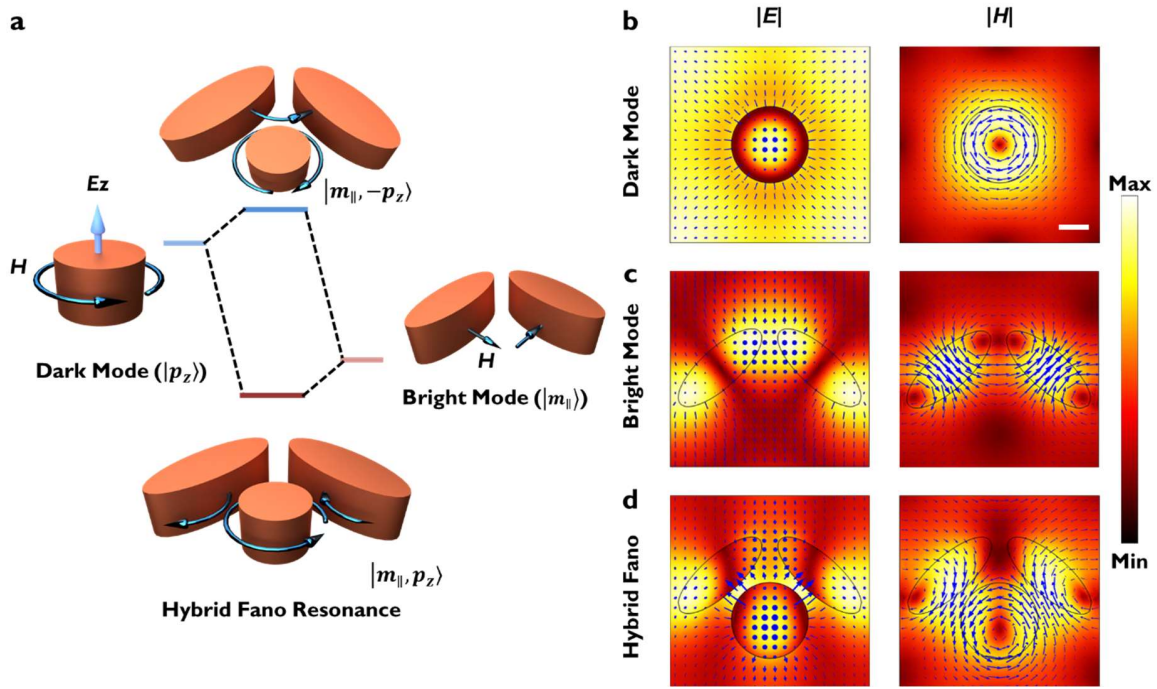

**Fig. 2 Design concept of dielectric Fano nanoantenna via mode hybridization. a** Schematic of hybridization between the dark mode ($|p_z\rangle$) of the disk structure and the bright mode ($|m_\parallel\rangle$) of paired ellipses. **b-d** Simulated spatial electric and magnetic field distribution at dark mode of disk, bright mode of the ellipsis pairs and hybrid Fano resonance mode. The scale bar denotes 50 nm.

Figure 2 presents the basic principle of mode hybridization for the designed dielectric Fano antenna. To analyze the mode mechanism, we consider the dielectric material as loss free Si with a refractive index of $n = 3.74$, thereby eliminating the influence of dispersion and Ohmic loss. As shown in Figure 2a, both the disk and elliptical pillar pairs exhibit transverse magnetic modes (*i.e.*, $H_z$ is negligible). The disk supports a non-radiative electric dipole oriented along *z*-axis at the wavelength of 561 nm, denoted as $|p_z\rangle$, generating an in-plane magnetic loop with the magnetic field ***H*** circulating in the *xy*-plane (Figure 2b). The ellipse pairs with a radiative quasi-BIC mode induce two in-plane magnetic dipoles $|m_\parallel\rangle$ at the wavelength of 651 nm (Figure 2c). The *x*-components of two magnetic dipoles $|m_\parallel\rangle$ are aligned in the same direction, thereby reinforcing each other along the *x*-axis; meanwhile, the *y*-components of magnetic dipoles are oriented in opposite directions, thus cancelling each other out. In other words, this quasi-BIC mode can be readily excited when the incident optical field is *y*-polarized with a dominating magnetic field being *x*-polarized.

When integrating ellipse pair and a disk in a single unit cell, two hybrid modes are generated at 550 and 671 nm, respectively, due to the coupling between $|p_z\rangle$ and $|m_\parallel\rangle$. To be more precise, the upper hybridized mode ($|m_\parallel, -p_z\rangle$) arise from the coupling between $|m_\parallel\rangle$ along +*x* axis and $|p_z\rangle$, where its corresponding electric and magnetic field distributions are shown in Figure S3. On the other hand, the lower one ($|m_\parallel, p_z\rangle$) is due to the interaction of $|m_\parallel\rangle$ and $-|p_z\rangle$ (see Figure 2d). Consequently, a narrow dip emerges in the reflection spectrum, as shown in Figure 1b, where this mode is able to achieve a strongly enhanced optical field at the nanogap region between disk and ellipse, especially at the plane of *z*=25 nm above substrate for strong Purcell enhancement (see the electric and magnetic field distributions at Figure S4). Moreover, we have also investigated the dependence of this hybridized mode on the nanogap size, which is a key parameter to engineer the ultimate field enhancement factor and Purcell effect (see Figure S5).

The photoluminescence peak was measured to be located at 657 nm, indicating the existence of negatively charged nitrogen-vacancy (NV$^-$), instead of NV$^0$ (see SI, Figure S6). Therefore, nanoarray with disk diameter of 120 nm is the most optimized value due to the highest electric field intensity enhancement and a good spectrum overlap between the sharp Fano resonance and the emission wavelength of NV centers as measured, where the detailed disk optimization is shown in Figure S7.

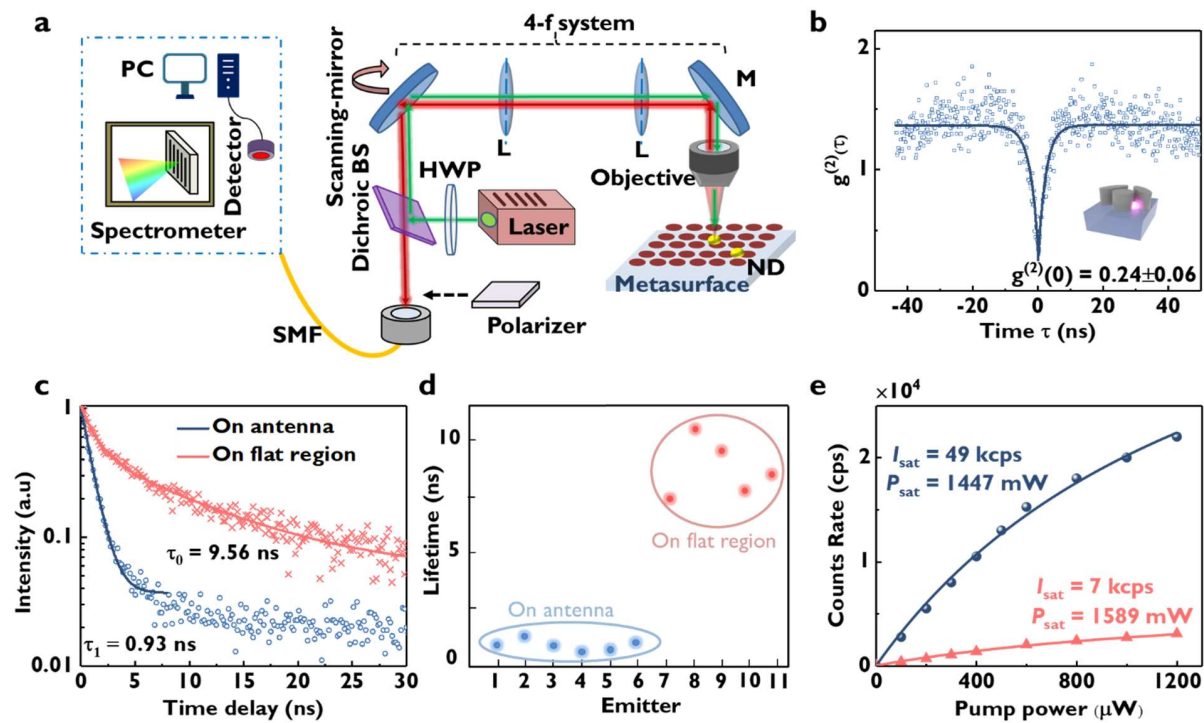

**Fig. 3 Optical characterization of quantum emitters integrated with Si Fano nanoantenna. a** Optical setup for the emission lifetime and second order photon correlation $g^{(2)}(\tau)$ measurements. HWP：half waveplate, L: lens, M: mirror, SMF: single mode fiber. **b** Measurement of the second-order autocorrelation, $g^{(2)}(\tau)$, of a representative NV center on the metasurface under 515 nm CW excitation with a power of 1.2 mW. The blue dots represent the raw data, while the solid blue line

is the fitting using the three-level model[37]. **c** Representative normalized fluorescence decay lifetimes for NV centers located on sapphire substrate (red) and Si Fano antenna (blue). The fitting for lifetime of emitters on the metasurfaces only used the first 50 points since the other data were at noise level. **d** Statistical comparison between the sets of NVs on flat region and antenna coupled NVs in terms of lifetime and $g^{(2)}(0)$. **e** Emission saturation dependencies for quantum emitters on pump laser power on the metasurface (blue) and on the flat region (red).

Figure 3a presents the home-built confocal microscope setup, which was able to measure the lifetime and second-order correlation function, $g^{(2)}(\tau)$, for the characterization of SPEs being coupled to Si Fano antenna. The excitation was performed by 515-nm laser, which is capable to operate either in continuous wave (CW) or pulsed modes (110 ps). PL intensity map measurement was first conducted to explore the distribution of the emitters (Figure S8). To verify the emission of single photons, we performed the $g^{(2)}(\tau)$ measurement for a representative single NV$^-$ center coupled to Fano resonance using a Hanbury-Brown and Twiss (HBT) system. The collected photon was directed to a fiber-based beam splitter and avalanche photodiodes (APDs) via single mode fiber. The correlation function shown in Figure 3c exhibits a distinctive dip at $\tau = 0$ with $g^{(2)}(0) = 0.24$, proving the antibunching behaviour of the emitted single photons.

Figure 3c depicts a comparison of the lifetime between the SPEs on Si metasurface and on flat substrate region. Notably, the emission lifetime of SPE on metasurface is significantly reduced to 0.93 ns (red curve); in comparison, the emitter at the flat substrate region has a lifetime of 9.56 ns, revealing a Purcell factor of ~10. The strong reduction of lifetime is a direct consequence of the higher density-of-optical-states in the nanogap between disks and the ellipses, which include contribution from the enhanced radiative emission (Purcell effect). This experimental result verifies the simulation investigation (Figure 1d) on the peak Purcell factor. The difference in simulated and observed Purcell enhancement is explained by the fact that emission linewidth of NV⁻ centers is larger rather than width of the resonance, which makes enhancement less efficient and thus reduces the effective Purcell factor [38]. Similar decay dynamics were observed from other 5 coupled and 5 uncoupled single photon and quantum emitters, and their extracted lifetime are summarized in Figure 3d and detailed in Table S1. These results confirm that the metasurface with Fano resonance could reduce the lifetime of the NV⁻ center emitter effectively. Additionally, we measured the emission photocounts as a function of the pump power shown in Figure 3e. The data were fitted with the equation $I(P) = I_{sat} \times P/(P + P_{sat})$ [39], where $P$ is the pump power, $I_{sat}$ and $P_{sat}$ represent fitting parameters that denote the saturation intensity and saturation power, respectively. From the measured pump dependence, we found that the enhancement of the emission rate constituted 7.5 times in the linear regime. The measured emission rate enhancement agrees well with the lifetime reduction, proving the absence of Ohmic dissipation loss in dielectric structures.

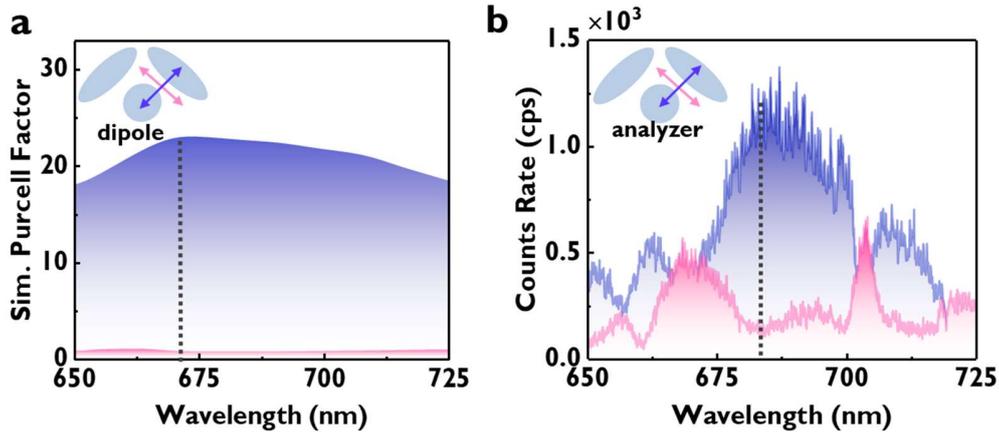

**Fig. 4 Optical characterization of the polarization characteristics of single photon emitters. a** Simulated Purcell factors of dipoles oriented at 45° (blue) and 135° (pink), respectively, when excited at the hot spot. **b** Measured counts rate of SPEs on metasurface with polarization selective analysers along various directions at wavelengths ranging from 650 to 725 nm.

We also explored the polarization properties of single photon emission. We simulated the Purcell factor under various dipole excitations (Figure 4a). It is observed that positioning a dipole at a 45-degree angle results in a Purcell factor as high as 23, indicating significant radiation enhancement. In contrast, setting the dipole at a 135-degree angle leads to a Purcell factor of only 0.95, demonstrating radiation suppression. Therefore, the degree of polarization (DOP) of single photon emission was calculated to be ~0.92. To verify the emission polarization experimentally, an analyser composed of a half-wave plate followed by a linear polarizer was put before collection single mode fiber and then sent to spectrometer. The collected photons exhibit the largest contrast with a ratio of 9 at the wavelength of 680 nm, which matches well with the simulation result (Figure 4b). These observations clearly demonstrate strongly polarized single photon emission.

**Conclusion**

In this paper, we have successfully designed a Fano dielectric nanoantenna to enhance the spontaneous emission rate of quantum emitters. We demonstrate the functionality of our platform by integrating it with NV$^-$ center in NDs and achieving an experimental Purcell enhancement factor of ~10 fold, confirmed by measurements of lifetime reduction and emission rate enhancement. Furthermore, the polarization contrast of SPEs induced by metasurface was demonstrated to be 9, providing a linearly polarized SPE. We believe that our work will contribute to the development of more efficient and scalable quantum light sources for quantum computation and communication systems.

**METHODS**

**Sample fabrication.** The process began by thinning the Si layer thickness on commercially purchased silicon-on-sapphire (SOS, Silicon Valley Microelectronics, Inc.) wafers from 300 to 240 nm, using chlorine ($Cl_2$)-based inductively-coupled-plasma reactive-ion etching (ICP RIE, Oxford Instruments Plasmalab System 100). Next, 6% HSQ was spin coated on the SOS sample at a speed of 5000 rpm for 60 s to form a layer of 70 nm thickness. After that, an array of ellipses with disks nanopatterns was fabricated on the SOS substrate using electron beam lithography (EBL), where the size of the patterning area was 50 × 50 μm$^2$. In the EBL process, an acceleration voltage of 100 keV and a beam current of 500 pA, with an exposure dosage of ~25 mC/cm$^2$ was used. The Si metasurface was then completed by etching down to sapphire substrate.

**Nitrogen-vacancy (NV⁻) centers in nanodiamond and sample dip-coating.** The NV centers in nanodiamonds with a diameter of 40 nm (Adamas Nanotechnologies) suspended in isopropanol with a concentration of 0.1 Millimoles/L. Before the deposition process, we carried out the surface treatment process via $O_2$ plasma using RIE (Oxford Plasmalab 80) with a pressure of 10 mTorr, a power of 50 watts and an $O_2$ flow rate of 50 sccm for 2 minutes. Subsequently, the sample was dip-coated inside the NV⁻ in nanodiamond solution at a rate of 0.1 mm/min within a nitrogen-filled glove box at room temperature [40].

**Numerical simulations.** The reflectance spectra were obtained using Lumerical 3D finite difference time-domain (3D-FDTD) software. The elliptical structures with disk were positioned in a unit cell measuring 327 nm × 324 nm × 1500 nm, with a 2.5 nm × 2.5 nm × 2.5 nm mesh surrounding them. This unit cell featured a perfectly matched layer (PML) along the $z$ direction and periodic boundary conditions along the $x$ and $y$ directions. The reflectance spectrum was recorded under normal incidence conditions using a monitor placed above a plane wave source. The incident light wave was polarized along either the $x$- or $y$-axis. The refractive index ($n$) and extinction coefficient ($k$) values for crystalline Si were obtained from ellipsometer measurements[29]. The Purcell factor was calculated using the software's built-in method, based on the ratio of dipole emission power to total emitted power. The domain was constrained by PML boundary conditions. A point dipole source was positioned within a 13 × 13 array of nanostructures, specifically at the location of the highest field enhancement in the center nanoellipse. The emission center was set at 680 nm with a 100 nm span, and the simulation results of overall Purcell factor were averaged from orthogonal dipole orientations along the x-, y-, and z-axes.

The electric and magnetic field distributions in Figure 2 were obtained with finite-element method in Electromagnetic Waves Frequency Domain module of COMSOL Multiphysics. All modal analysis were investigated using eigenfrequency solver. In all cases, periodic boundaries were applied to x and y axis, while scattering boundaries are applied on both +z and -z boundaries. A Tetrahedral mesh with local refinement around fine structures was employed. The refractive index of sapphire was set as 1.76. The structural parameters were set based on the SEM of as-fabricated structures. Due to some dimensional tolerance in SEM measurements (about ±5 nm), the parameters were adjusted within a reasonable range according to the comparison between simulated and measured spectra.

**Lifetime measurement and $g^{(2)}(\tau)$ measurement.** The measurements were performed at the home-build confocal microscope setup, where the 515 nm pump laser (Omicron) was scanned by steering mirror (Newport) and projected to the sample by 4f lens system and 100x microscope objective (Olympus, NA = 0.9). Lifetime measurements were conducted using 515 nm laser set in the pulsed mode (110 ps pulse duration, 2 MHz repetition rate), and emitted photons were detected by a single-photon avalanche diode (SPAD) (Excilatas). The correlations between the emission detected by SPAD and the laser pulses were measured using a time-to-digital converter (QuTools quTau), with a resolution of 81 ps. Spectral filtration was achieved using a Semrock 680/42 filter. The fluorescence decay curve was fitted to a single exponential model to extract the lifetime. For $g^{(2)}(\tau)$ measurements, the pump laser at 515 nm was set in the CW mode. The emission was selected by Semrock filter 675/150 and split between two SPADs by the fiber beamsplitter (Thorlabs). The same time-to-digital converter (QuTools quTau) recorded photon arrival time differences to calculate $g^{(2)}(\tau)$, with $g^{(2)}(0) < 0.5$, indicating single-photon emission. The spectrometer measurements were performed by sending the collected emission into the spectrometer a spectrometer (Princeton Instruments, IsoPlane160, 0.13 nm max resolution). Calibration was performed using known standards, and error analysis accounted for incertainty sources such as photon counting statistics and detector noise.

## ASSOCIATED CONTENTS

**Supporting Information**

## AUTHOR INFORMATION

**Corresponding Author**

*E-mail: dongz@imre.a-star.edu.sg

*Email: Victor_Leong@imre.a-star.edu.sg

*Email: chengwei.qiu@nus.edu.sg

**ORCID**

Shu An: 0009-0002-2878-7960

Dmitry Kalashnikov: 0000-0002-8954-6178

Wenqiao Shi: 0009-0001-2870-2413

Di Zhu: 0000-0003-0210-1860

Weibo Gao: 0000-0003-3971-621X

Cheng-Wei Qiu: 0000-0002-6605-500X

Zhaogang Dong: 0000-0002-0929-7723

**Author contributions**

S.A., Z.D., C.-W.Q., D.K. and V.L. conceived the concept, designed the experiments. S.A. did the electron beam lithography, developed the fabrication processes, and performed the simulations and optical reflectance measurements. D.K. measured the lifetimes and second-order correlation function of the emitters. W.S. performed the simulation analysis. Z.M. prepared the ND suspension and optimized the dip coating process to deposit the NDs. A.B.C. helped with the SEM image and interpretation. Y.L. provided support on the mechanism analysis. J.W., D.Z. and W.G. participated in discussions and provided suggestions to improve the work. Z.D., C.-W.Q. and V.L. supervised the project. All authors analyzed the data, read and corrected the manuscript before the submission.

**Notes**

**ACKNOWLEDGMENTS**

This work is supported by the funding support of Agency for Science, Technology and Research (A*STAR) with the Project No. C230917005. In addition, Z.D. also would like to acknowledge the funding support from A*STAR under its Career Development Award grant (Project No. C210112019), MTC IRG (Project No. M21K2c0116 and M22K2c0088), the Quantum Engineering Programme 2.0 (Award No. NRF2021-QEP2-03-P09), and National Research Foundation via Grant No. NRF-CRP30-2023-0003.